\documentclass{osa-article}
\journal{oe}
\articletype{Research Article}

\usepackage{subfigure}
\usepackage{multirow}
\usepackage{indentfirst}
\usepackage{color, colortbl}
\usepackage{makecell}
\graphicspath{{images/}}

\begin{document}


\title{Graded Index Couplers for Next Generation Chip-to-Chip and Fiber-to-Chip Photonic Packaging}

\author{Drew Weninger\authormark{1, 4}, Christian Duessel\authormark{1,4}, Samuel Serna\authormark{3}, Lionel Kimerling\authormark{1,4}, and Anuradha Agarwal\authormark{2,4}}

\address{\authormark{1} Materials Science and Engineering Department\\
\authormark{2} Materials Research Laboratory\\
\authormark{3}Bridgewater State University, Physics Department, 131 Summer St, Bridgewater, MA 02324, USA\\
\authormark{4}Massachusetts Institute of Technology, 77 Massachusetts Avenue, Cambridge, MA 02139, USA}

\email{\authormark{*}drewski@mit.edu}


\begin{abstract}
The transition towards designs which co-package electronic and photonic die together in data center switch packages has created a scaling path to Petabyte per second (Pbps) input/output (I/O) in such systems. In a co-packaged design, the scaling of bandwidth, cost, and energy will be governed by the number of optical I/O channels and the data rate per channel. While optical communication provide an opportunity to exploit wavelength division multiplexing (WDM) to scale data rate, the limited 127 $\mu$m pitch of V-groove based single mode fiber arrays and the use of active alignment and bonding for their packaging present challenges to scaling the number of optical channels. Flip-chip optical couplers which allow for low loss, broadband operation and automated passive assembly represent a solution for continued scaling. In this paper, we propose a novel scheme to vertically couple between silicon based waveguides on separate chips using graded index (GRIN) couplers in combination with an evanescent coupler. Simulation results using a 3D Finite-Difference Time-Domain (FDTD) solver are presented, demonstrating coupling losses below 0.27 dB for a chip-to-chip gap of 11 $\mu$m; 1-dB vertical and lateral alignment tolerances of approximately 2.38 $\mu$m and $\pm$ 2.24 $\mu$m, respectively; and a greater than 360 nm 1-dB bandwidth. These results demonstrate the potential of our coupler as a universal interface in future co-packaged optics systems.
\end{abstract}


\section{Introduction}
\par Silicon photonic integrated circuits (Si-PICs) have seen significant advancement over the past 40 years, highlighted by their rise to dominance as data center interconnects \cite{2024shekhar} and the novel application of Si-PICs as biochemical sensors \cite{2010iqbal}, in LIDAR \cite{2022zhang}, photonic switching \cite{2019seok}, photonic computing \cite{2021shastri}, and even chip based 3D printing \cite{2024corsetti}. However, the scaling of optical input/output (I/O) remains a significant manufacturing challenge, underscored photonic packaging, assembly, and testing occupying 70-80\% of the total cost of Si-PIC manufacturing \cite{2022ranno}. One of the primary barriers to optical I/O scaling is the use of active alignment and bonding of V-groove based single mode fiber (SMF) arrays to Si-PICs using UV-curable epoxies, increasing cost and limiting throughput. Not only are current optical packaging methods costly, but they are severely limited in terms of pitch - SMF arrays operating near datacom (1310 nm) or telecom (1550 nm) wavelengths have minimum pitches of 127 $\mu$m, meaning a maximum density of only 8 fibers/mm is possible. The assembly challenge is especially pertinent to co-packaged optics (CPO) systems, which may require $>$ 10$^3$ SMFs to scale to the Pbps package I/O data rates by 2035 as required by current CPO trends \cite{2021minkenburg}.

\par
\begin{figure}[htbp]
\centering
\subfigure[\label{c2c_overview_perspective}]{\includegraphics[scale=0.35]{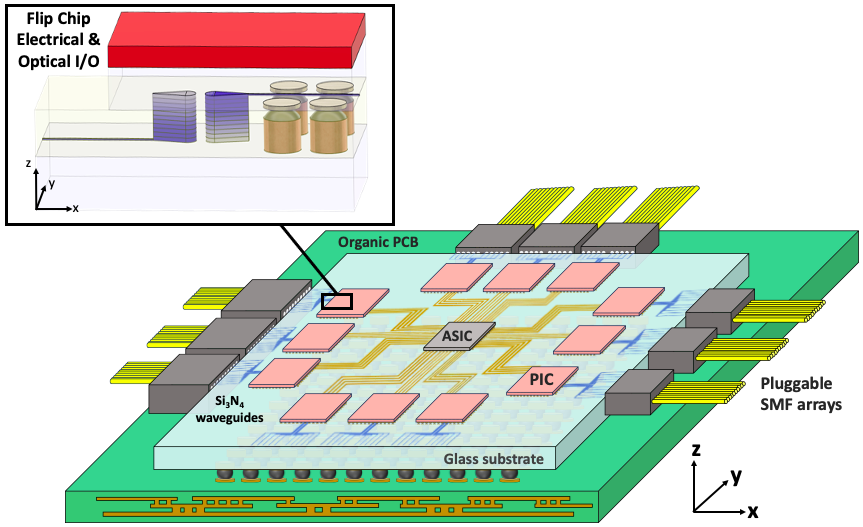}}
\subfigure[\label{c2c_overview_cross_section}]{\includegraphics[scale=0.35]{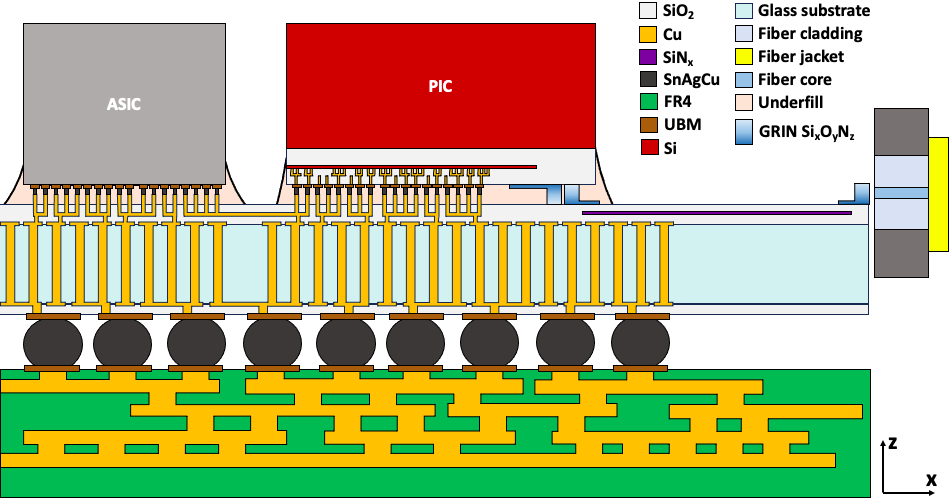}}
\caption{An example of the GRIN coupler deployed in a co-packaged optics system. In \subref{c2c_overview_perspective}, a 3D view shows light entering the edge of a glass package substrate using a GRIN fiber-to-chip coupler and being transferred to a Si-PIC using a GRIN chip-to-chip coupler. In \subref{c2c_overview_cross_section}, a 2D cross section shows the simultaneous optical and electrical fanout occurring between the ASIC, a single Si-PIC, and a single fiber array which is enabled by the GRIN coupler. The images above were adapted from \cite{2024weninger} to add the GRIN lens element.}
\label{fig:overview}
\end{figure}

 In light of these packaging and assembly related problems, we have developed and patented \cite{2024weninger_patent} an optical coupling solution based on integrated GRIN lenses and evanescent couplers which can be used for flip-chip (i.e. chip-to-chip) connections, analogous to the electrical solder joint. Moreover, this optical coupling solution can be simultaneously used for fiber-to-chip coupling and packaged alongside electrical back-end-of-line (BEOL) interconnects. An example of how this can be incorporated in a CPO system layout is shown in Fig. \ref{fig:overview}.  

\par This study focuses on the design and simulation of the GRIN coupler in the context of fiber-to-chip and chip-to-chip coupling. The simulations evaluated the coupling loss, 1-dB and 3-dB misalignment tolerances, 1-dB wavelength tolerance, and the fabrication tolerances of such a design. Moreover, Section \ref{sec:GRIN} describes the operational principle and novelty compared to prior designs, Section \ref{sec:methods} depicts the systems constraints and simulation setup, Section \ref{sec:results} presents the collected data, and Section \ref{sec:discussion} provides takeaways and a comparison to similar designs.

\subsection{Graded index (GRIN) optical couplers}\label{sec:GRIN}
\par
\begin{figure}[!t]
\centering
\subfigure[\label{f2c_perspective}]{\includegraphics[scale=0.2]{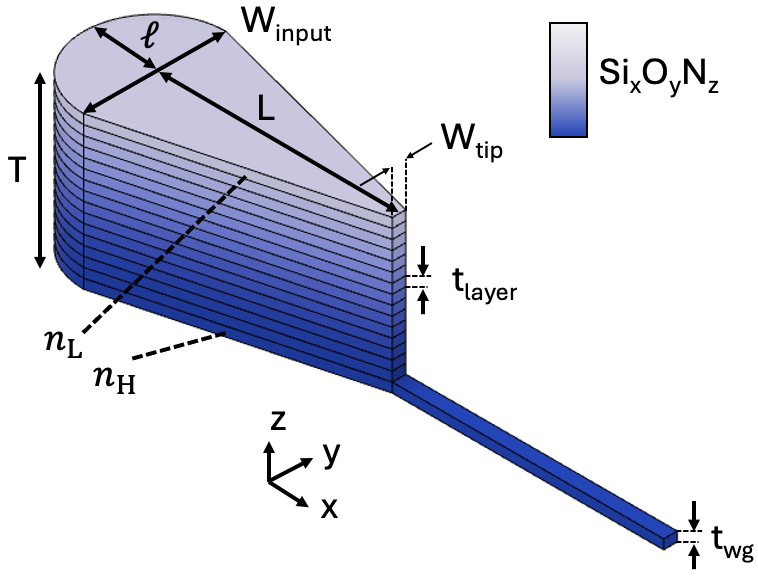}}
\subfigure[\label{c2c_schematic}]{\includegraphics[scale=0.4]{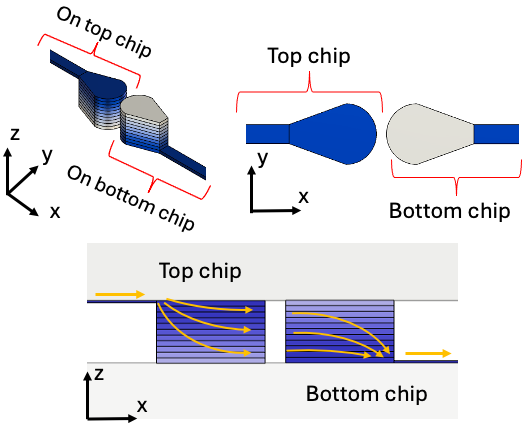}}
\subfigure[\label{low_index_plot}]{\includegraphics[scale=0.3]{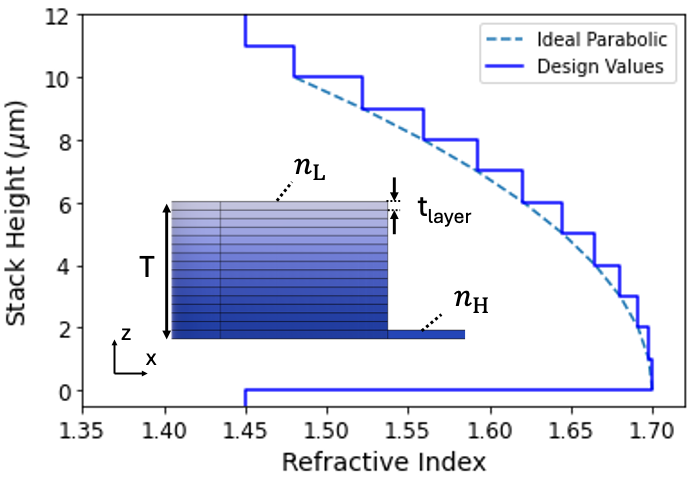}}
\caption{Schematics showing the GRIN coupler. In \subref{f2c_perspective}, a 3D view showing the basic building block of the GRIN coupler used for fiber-to-chip coupling. In \subref{c2c_schematic}, a 3D view, top down view, and cross sectional view of how the GRIN coupler can be used for chip-to-chip coupling. In \subref{low_index_plot} a plot of the refractive index of the GRIN lens as a function of thickness is shown for the ideal parabolic profile and the step profile.}
\label{device_schematic}
\end{figure}

 The basis for our GRIN coupler design involves using silicon oxynitride (Si$_x$O$_y$N$_z$, abbreviated SiON for the remainder of this report) layers with a GRIN profile to vertically focus, and a lens and taper combination to horizontally focus the incoming optical mode, as shown in Fig. \ref{device_schematic}. Our design builds on prior efforts from \cite{2006nguyen, 2007sun} and is similar to the variations found in \cite{2010loh, 2016lim, 2024xu}. However, our design is distinct from these in several ways. First, it is intended for coupling between two separate chips or from interposer to chip (inter-chip coupling) in addition to fiber-to-chip coupling. By using a GRIN coupler for chip-to-chip coupling, we can increase the allowable vertical gap between chips by an order of magnitude compared to other types of chip-to-chip couplers such as evanescent, grating, or edge based methods. An inter-chip gap $>$ 10 $\mu$m will allow for the integration of electrical connections such as Cu $\mu$-pillar bumps without the need for fabricating them in a trench. Moreover, this can be accomplished without sacrificing the ability to process GRIN couplers in parallel at the wafer scale, unlike couplers reliant on TPP processes which are fabricated serially. The goal of these couplers will be to increase lateral 1-dB alignment tolerance beyond the capability of the automated pick-and-place die bonders found in high volume manufacturing so that passive assembly can be utilized. Widening alignment tolerances to enable passive assembly of optical and electrical components directly equates to faster assembly speeds and lower costs.

\par The second novel aspect of our design is the addition of an evanescent coupling element after the GRIN lens, to allow for coupling directly to standard 220 nm SiN or 220 nm SOI waveguides, or other desirable material platforms such as InP or LiNbO$_3$. By integrating an evanescent coupling element, not only can the input and output waveguide material be customized, but the GRIN lens can be deposited in the back-end-of-line (BEOL) using only low-temperature plasma enhanced chemical vapor deposition (PECVD) processes ($<$ 350\textdegree C) with no additional annealing. In addition, with minimum SiON feature sizes as large as 500 nm, the GRIN coupler can be patterned using more mature technology nodes such as i-line (365 nm) photolithography with standard reactive ion etching (RIE). The aforementioned processes also maintain compatibility with microelectronic foundry tools used in CMOS process flows, leading to a decrease in fabrication complexity, and thus cost. Finally, by providing an interface for both chip-to-chip and fiber-to-chip coupling, one which is material agnostic to the input and output waveguides and can be added in the BEOL, the GRIN coupler can be seen as a universal interface. The remainder of this study will discuss how the simulations were setup and will show the final results demonstrating high coupling efficiency with widened 1-dB alignment tolerances.

\section{Materials and methods}\label{sec:methods}
\subsection{Material platform constraints for Si$_x$O$_y$N$_z$ GRIN coupler}
\par The design of the GRIN coupler, in theory, required the optimization of many variables; however, in practice this process was simplified by design constraints. First, the design methodology involved selecting the mode field diameter of the incoming beam (MFD$_i$), and the refractive index ($n_{\text{wg}}$) and thickness ($t_{\text{wg}}$) of the waveguides being coupled into. A flat facet, SMF-28 fiber with a 10.4 $\mu$m MFD was assumed, and waveguides with $n_{\text{wg}}$ = 1.7 and $t_{\text{wg}}$ = 1 $\mu$m were chosen as the output from the GRIN coupler based on prior data for depositing and patterning NH$_3$ free SiON films using PECVD \cite{2007sun}. We would like to emphasize that selection of these three parameters (MFD$_i$, $n_{\text{wg}}$, $t_{\text{wg}}$) was arbitrary - the operation of the GRIN coupler and the optimization methodology described in this section can be applied to any new set of selections. With these selections, all remaining parameters including the GRIN thickness ($T$), GRIN width (W$_{\text{input}}$), number of layers ($N$), tip width (W$_{\text{tip}}$), waveguide width (W$_{\text{wg}}$), and layer refractive indices ($n_{\text{i}}$ to $n_{\text{f}}$) were constrained, except the lens radius ($r$) and taper length ($L$), as outlined in Table \ref{GRINparameters}. 

\par In particular, $T$ and W$_{\text{input}}$ were constrained to be approximately equal to MFD$_i$ while W$_{\text{wg}}$ and W$_{\text{tip}}$ were constrained to the maximum waveguide width possible for the selections of $n_{\text{wg}}$ and $t_{\text{wg}}$. To determine this maximum waveguide width, 2D Finite-Difference Eigenmode (FDE) simulations were executed which swept waveguide width assuming a cross section described by $n_{\text{wg}}$ and $t_{\text{wg}}$, and surrounded by material with a refractive index equal to SiO$_2$ ($n_{SiO2}$ = 1.444 at 1550 nm). Furthermore, $N$ was constrained by $t_{\text{wg}}$ and $T$ such that the thickness of each layer in the GRIN was equal to $t_{\text{wg}}$. This constraint provides a reasonable minimum number of layers that should be used in the GRIN coupler which can always be increased for more advanced and optimized deposition processes. Finally, the coupler refractive index profile was parabolic from $n_{\text{L}}$ to $n_{\text{H}}$, as shown in Fig. \ref{low_index_plot}, using the following equation:

\begin{align}
\label{equation1} n(z) &= n_H\left(1-\frac{\alpha^2 z^2}{2}\right)\\
\label{equation2} \alpha^2 &= \frac{2\left(1-\frac{n_L}{n_H}\right)}{T^2}
\end{align}

The parabolic refractive index profile is necessary in order create the lensing effect predicted by Fermat's Principle and the ray equation \cite{2007saleh}. With these selections and constraints, $r$ and $L$ were swept over using 3D-FDTD simulations in Ansys Lumerical to optimize the design for maximum fiber-to-chip and chip-to-chip coupling efficiency. In order to properly form a lens with a given radius on the GRIN coupler, an equation for the lens surface needed to be developed. The lens surface is constrained such that it must intersect exactly at the edge of the GRIN at the points (0,$\frac{\text{W}_{\text{input}}}{2}$) and (0,$\frac{-\text{W}_{\text{input}}}{2}$) according to the coordinate system in Fig. \ref{f2c_perspective} (assuming ($x$,$y$) = (0,0) is at the center of the input of the GRIN coupler). For a circle of a given radius in the $x$-$y$ plane with a center at ($x_1$, $y_1$) governed by the equation $r^2$ =  $(x-x_1)^2$ + $(y-y_1)^2$, we can thus solve for the location of the center of the circle as $y_1$ = 0 and $x_1 = \sqrt{r^2 - \left(\frac{\text{W}_{\text{input}}}{2}\right)^2}$. The equation of the lens is then:

\begin{equation} 
\label{equation3} x = \sqrt{r^2 - \left(\frac{\text{W}_{\text{input}}}{2}\right)^2} - \sqrt{r^2 - y^2}
\end{equation}
where $x_1 \geq 0$ and $x \leq 0$. The distance $\ell$ as described by Fig. \ref{f2c_perspective} is obtained by inserting $y$ = 0 into equation \ref{equation3} and multiplying by a negative (because $x \leq 0$)  such that:
\begin{equation} 
\label{equation4} \ell = r - \sqrt{r^2 - \left(\frac{\text{W}_{\text{input}}}{2}\right)^2}
\end{equation}

From these relationships, it is clear that the minimum $r$ value that can be used is $\frac{\text{W}_{\text{input}}}{2}$ and as $r \rightarrow \infty$ that $\ell \rightarrow 0$ and the GRIN becomes flat faceted. This fact was used to establish bounds for sweeping the value of $r$ during simulations to be 5.5-100 $\mu$m. Similarly, the focal length of an asymmetric GRIN lens with a parabolic refractive index profile can be approximated using the period of the ray oscillation, $\frac{2\pi}{\alpha}$, as described in \cite{2007saleh}. The focal length of the asymmetric GRIN lens is then one quarter of the period (i.e. $\frac{\pi}{2\alpha}$). For the refractive index profile described by Table \ref{GRINparameters}, this indicates $\alpha \approx 0.051 \ \mu m^{-2}$ and the focal length for a flat faceted GRIN is approximately 30 $\mu$m. This fact was used to establish bounds for sweeping the value of $L$ during simulations to be 22-32 $\mu$m, centered slightly below 30 $\mu$m, since the lens section also contributes to vertical focusing. 

\begin{table}[!b]
\captionsetup{font=normalsize}
\centering
\renewcommand{\arraystretch}{1.2}
\caption{Summary of the design selections and constraints for parameter optimization.} 
\footnotesize
\begin{tabular*}{0.97\textwidth}{c|c|c|c|c|c|c|c|c|c|c}
\cline{2-11}
& \multicolumn{3}{c|}{\textbf{Selections}} & \multicolumn{7}{c}{\textbf{Constraints}}\\
\cline{1-11}
\textbf{Parameter} & MFD$_i$ & $n_{\text{wg}}$ & $t_{\text{wg}}$ & $T$ & W$_{\text{input}}$ & $N$ & 
W$_{\text{tip}}$ & W$_{\text{wg}}$ & $n_{\text{i}}$ & $n_{\text{f}}$\\
\cline{1-11}
\textbf{Value}& 10.4 $\mu$m & 1.7 & 1 $\mu$m & 11 $\mu$m & 11 $\mu$m & 11 & 1.2 $\mu$m & 1.2 $\mu$m & 1.48 & 1.7\\
\cline{1-11}
\end{tabular*}
\label{GRINparameters}
\end{table}

\subsection{Setup for 3D-FDTD GRIN coupler simulations}

\par The performance metrics of interest in this study are coupling loss (in units of dB), 1-dB and 3-dB lateral alignment tolerance (in units of $\mu$m), 1-dB and 3-dB vertical alignment tolerance (in units of $\mu$m), and 1-dB bandwidth (in units of nm).

\par For the 3D-FDTD simulation setup, meshes which were one-fifth the feature size were used in order to optimize simulation time. This coarse mesh was validated by the fact that all device dimensions were $>$ 1 $\mu$m scale. The background mesh for the solver was set to Lumerical's preset auto-non uniform mesh with a mesh accuracy of ``2'' (this equates to approximately a 100 nm background mesh for a wavelength of 1.55 $\mu$m). Moreover, perfectly-matched layer (PML) boundary conditions, which assume absorption of incident electric field, were used to account for loss due to out of plane scattering of light. For fiber-to-chip coupling, a Gaussian source with a waist radius of 5.2 $\mu$m and wavelength of 1.55 $\mu$m was placed 0.5 $\mu$m in front of the edge of the lens of the GRIN coupler. Transmission was measured through an output monitor 4 $\mu$m by 4 $\mu$m in size located 7 $\mu$m past the tip of the GRIN coupler. For the 1-dB bandwidth calculations, the source wavelength spectrum was edited to range from 1.28 $\mu$m to 1.64 $\mu$m to reflect the wavelength regime used for tele- and data-communications. The refractive index of SiON in the wavelength regime simulated was assumed to be constant for all layers, an assumption supported by the infrared ellipsometry data presented in Section \ref{sec:fab} and in literature \cite{1987bossi}. Because the Gaussian source does not use guided modes to create the beam, the electric and magnetic field angles of the incident beam were always perpendicular to the direction of propagation (i.e. a fundamental transverse electro-magnetic (TEM) mode).

\par For chip-to-chip coupling simulations, a monochromatic TE mode source with a wavelength of 1.55 $\mu$m was inserted in the input waveguide 6 $\mu$m prior to the tip of the GRIN lens. Transmission was measured through an output monitor 4 $\mu$m by 4 $\mu$m in size located 7 $\mu$m past the tip of the GRIN coupler. Similar to the fiber-to-chip coupling simulations, the 1-dB bandwidth calculations changed the source wavelength spectrum to extend from 1.28 $\mu$m to 1.64 $\mu$m with a wavelength step of 5 nm. In the chip-to-chip coupling simulations, the polarization of the mode source was also shifted from TE mode to TM mode to determine polarization dependent losses. 
 
\par To determine the 1-dB and 3-dB lateral ($y$), vertical ($z$), and longitudinal ($x$) alignment tolerances, sweeps were performed where the $y$, $z$ ,and $x$ misalignment of the GRIN coupler on the lower chip was varied from 0 to 3.5 $\mu$m. The $y$ and $z$ sweeps were performed assuming a 1 $\mu$m spacing in the $x$ direction between the surface of the GRIN lens on either die. In terms of $z$ alignment, the optimal position occurs when the bottom die GRIN coupler is in contact with the top die surface as shown in Fig. \ref{c2c_schematic}, and vice versa. The vertical distance from the top die substrate to the bottom die substrate is equal to $T$ (11 $\mu$m). Thus, a positive vertical misalignment indicates a $z$ offset between the two chips in addition to $T$. 

\par Simulations to determine the fabrication tolerance of the layer refractive index profile and thickness for chip-to-chip coupling were also executed. This was done using 2D-FDTD simulations where the simulation was performed across the $y$ = 0 plane. In the simulation, a sweep was performed for 0 - 1.5\% maximum error in the refractive index of a given layer. To do this, a maximum percent error was assigned ($\delta$), and then $N_s$ simulations were performed in which the refractive index of each layer of the GRIN lens was simultaneously perturbed by a different percentage randomly varying from -$\delta$ to $\delta$ (i.e. $n_{\text{layer}} = n_{\text{layer}} + Rn_{\text{layer}}\delta$ where $R$ is a random number between -1 and 1). For our simulations, an $N_s$ of 100 was used. An identical process was followed for random fluctuations in the thickness of each layer, except the sweep was performed from 0 - 20\% maximum error. Because the focal length changes slightly for the 2D versus 3D simulations (as the 2D simulations assume an infinitely wide GRIN lens), the focal length was re-optimized by performing a length sweep prior to performing any layer fluctuation simulations. Note that these simulations do not take into account the impact of random fluctuations in index or thickness on the horizontal confinement of the mode, only the vertical confinement. However, because the primary impact of a small change in index or thickness is on the vertical confinement through equations \ref{equation1} and \ref{equation2}, these 2D simulations are valid to determine an approximate fabrication tolerance. 

\par Finally, simulations to determine the allowable variation in the underfill refractive index were conducted. This was done by varying the underfill index from that of air ($n_{\text{air}}$ = 1.0) to 1.6. During these simulations, the underfill is assumed to fill the entire gap between the two die. The results of these simulations will be presented in the following section. 


\subsection{Setup for 3D-EME evanescent coupler simulations}
\par To simulate the loss through the evanescent coupler connecting the SiON output waveguide from the GRIN coupler to the final silicon based waveguide, the 3D Eigenmode Expansion (EME) solver was used. The evanescent coupler involved using a double taper design and optimization process developed previously in \cite{2023weninger} and fabricated in \cite{2024weninger}, except in this case it was used for connecting SiON to SOI and SiON to SiN$_x$ instead of SOI to SiN$_x$. A top down view showing the double taper evanescent coupler can be found in Fig. \ref{evanescent_schematic} which uses the same labeling convention as \cite{2023weninger}. The optimization process involved running a single 3D-EME simulation for a set of W$_{\text{SiON, i}}$, W$_{\text{tip, SiON}}$, W$_{\text{Si, i}}$, W$_{\text{tip, Si}}$ values where W$_{\text{Si, i}}$ and W$_{\text{tip, Si}}$ could indicate SOI or SiN$_x$. Each simulation swept across taper length ($L_{\text{taper}}$) up to 100 $\mu$m to determine at what length an adiabatic transition is formed. These simulations were performed for a matrix of W$_{\text{SiON, i}}$ and W$_{\text{tip, SiON}}$ values ranging from 0.6-1.2 $\mu$m and 400-600 nm, respectively. A matrix of W$_{\text{Si, i}}$ and W$_{\text{tip, Si}}$ values ranging from 0.23-0.44 $\mu$m and 100-200 nm were used for SOI and values ranging from 0.55-1 $\mu$m and 100-200 nm were used for SiN$_x$, respectively. The vertical gap between the tapers was assumed to be 200 nm and filled with SiO$_2$. This assumption is based on prior fabrication processes used to build inter-layer evanescent couplers which involved SiO$_2$ cladding deposition followed by chemical mechanical polishing (CMP) for planarization \cite{2008sun}.



\section{GRIN coupler simulation results}
\label{sec:results}
\subsection{GRIN fiber-to-chip coupler simulation results}
\label{sec:f2c_results}

\par
\begin{figure}[!t]
\centering
\subfigure[\label{f2c_sweep_low_index}]{\includegraphics[scale=0.28]{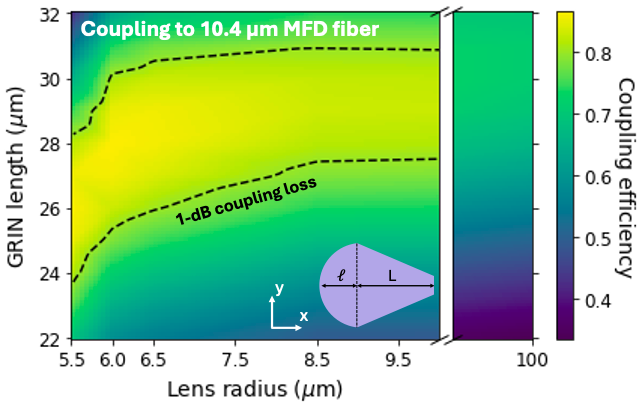}}
\subfigure[\label{f2c_misalign_sweep_low_index}]{\includegraphics[scale=0.32]{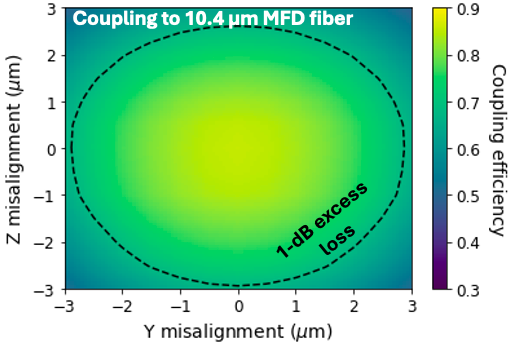}}
\subfigure[\label{f2c_mode_linear}]{\includegraphics[scale=0.3]{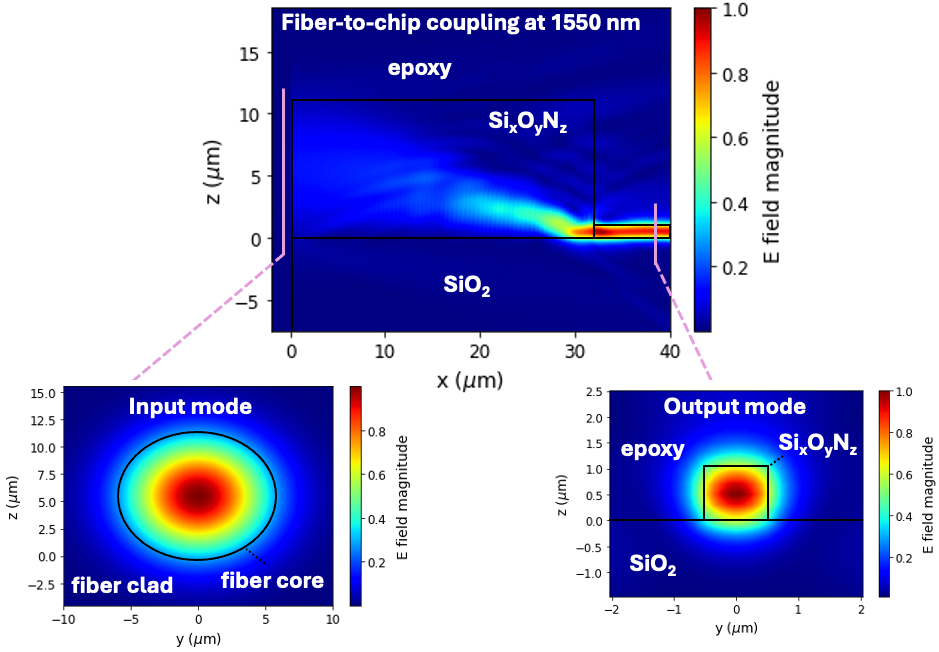}}
\caption{Simulations showing the GRIN coupler performance for fiber-to-chip coupling. In \subref{f2c_sweep_low_index}, coupling efficiency is shown for a 2D sweep of lens radius $r$ and GRIN length $L$. The dotted contour line represents 1 dB of total coupling loss. In \subref{f2c_misalign_sweep_low_index}, coupling efficiency is shown for lateral and vertical misalignments. The dotted contour line represents 1 dB of additional loss from the position of maximum coupling. In \subref{f2c_mode_linear} the transition of the optical mode from an MFD of 10.4 $\mu$m in the SMF to approximately 1.42 $\mu$m in the SiON waveguide is shown.}
\label{f2c_performance}
\end{figure}

 A plot showing the results for coupling efficiency at 1550 nm wavelength as a function of $r$ and $L$ can be found in Fig. \ref{f2c_sweep_low_index}. The creation of this plot involved linear interpolation of the simulated data to provide a smoother image and clearly show the dependencies on $L$ and $r$. Note that the reported values for coupling loss in this study were still taken from the original simulated values, while the interpolated values provided a useful means to determine approximate alignment tolerance.  From the data, a peak coupling efficiency of 86.6\% (0.62 dB loss) was observed for $r$ = 5.5 $\mu$m and $L$ = 26 $\mu$m. The allowable fabrication tolerance before 1 dB of total coupling loss was observed was approximately 2 $\mu$m for $L$ and $r$ based on the contour lines in Fig. \ref{f2c_sweep_low_index}. The coupling efficiency as a function of vertical and lateral misalignment can be found in Fig. \ref{f2c_misalign_sweep_low_index}, demonstrating a 1-dB lateral alignment tolerance of $\pm$ 2.88 $\mu$m. The vertical alignment tolerance depends on the direction of misalignment, since the GRIN lens is asymmetric. The 1-dB vertical alignment tolerance when the fiber is shifted in the negative $z$ direction is $\pm$ 2.99 $\mu$m and in the positive $z$ direction it is $\pm$ 2.58 $\mu$m.

\par A plot showing the modal transformation from the fiber core to the output SiON waveguide at 1550 nm for the fundamental TEM mode can be found in Fig. \ref{f2c_mode_linear}. The images detail how the 10.4 $\mu$m MFD from the SMF transitions using the GRIN lens to approximately 1.42 $\mu$m in the single mode waveguide. The modal images also involved linear interpolation to smooth the relatively coarse simulation output. Note that the wavelength dependency of the fiber-to-chip coupler can be found later in Fig. \ref{c2c_wvl_pol_low_index}, demonstrating a 1-dB bandwidth of over 360 nm from 1280 to 1640 nm.


\subsection{GRIN chip-to-chip coupler simulation results}
\label{sec:c2c_results}
\par
\begin{figure}[!t]
\centering
\subfigure[\label{c2c_sweep_low_index}]{\includegraphics[scale=0.28]{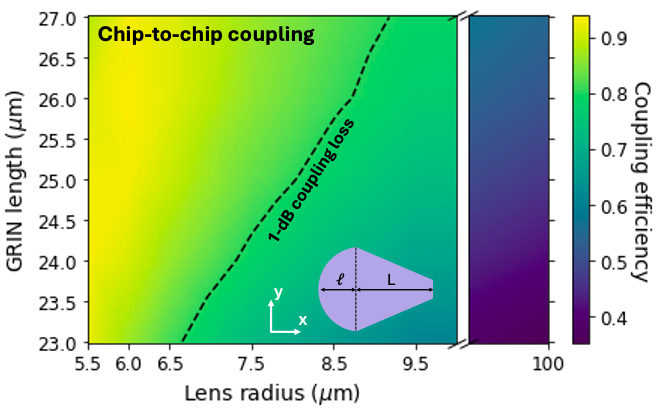}}
\subfigure[\label{c2c_misalignment_high_index}]{\includegraphics[scale=0.315]{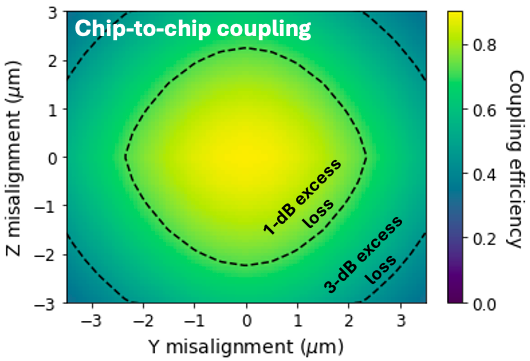}}
\subfigure[\label{c2c_mode_linear}]{\includegraphics[scale=0.32]{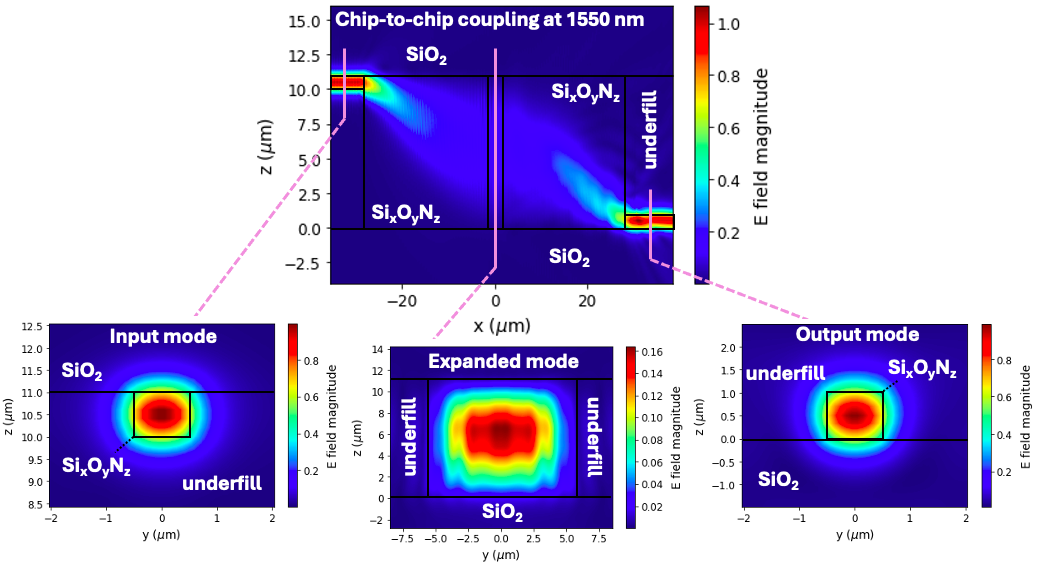}}
\caption{Simulations showing the GRIN coupler performance for chip-to-chip coupling. In \subref{c2c_sweep_low_index}, coupling efficiency is shown for a 2D sweep of lens radius $r$ and GRIN length $L$. The dotted contour line represents 1 dB of total coupling loss. In \subref{c2c_misalignment_high_index}, coupling efficiency is shown for lateral and vertical misalignments. The dotted contour lines represent 1 dB and 3 dB of additional loss from the position of maximum coupling. In \subref{f2c_mode_linear} the transition of the optical mode from a SiON waveguide on one chip to a SiON waveguide on a separate chip across an 11 $\mu$m chip-to-chip gap is shown.}
\label{c2c_performance}
\end{figure}

 A separate sweep of $r$ and $L$ was performed for the chip-to-chip coupler, such that the optimal $r$ and $L$ were 6 $\mu$m and 26 $\mu$m for a peak coupling efficiency of 93.9\% (0.27 dB loss) as shown in Fig. \ref{c2c_sweep_low_index}. The image was created using linear interpolation in an identical fashion to the fiber-to-chip simulations shown in Fig. \ref{f2c_sweep_low_index}. The data shows an exceptionally large fabrication tolerance of $> \pm$ 1 $\mu$m from the optimized design values for both $r$ and $L$. Likewise, the 1-dB lateral and vertical alignment tolerances were determined to be $\pm$ 2.38 $\mu$m and $\pm$ 2.24 $\mu$m, respectively, using the data in Fig. \ref{c2c_misalignment_high_index} which shows coupling efficiency as a function of $y$ and $z$ misalignment (after linear interpolation). Based on this data, the 3-dB lateral alignment tolerance was $\pm$ 3.81 $\mu$m and the vertical tolerance was $\pm$ 3.75 $\mu$m. 

\par Using the optimized $r$ and $L$, a plot showing the transformation of the fundamental TE mode during chip-to-chip coupling at 1550 nm wavelength, represented using the electric field magnitude on a linear scale, can be found in Fig. \ref{c2c_mode_linear}.  The (linearly interpolated) modal images demonstrate the expansion and focusing within the GRIN lenses which resulted in successful coupling over an 11 $\mu$m vertical gap. In order to provide more accurate cross sectional modal plots, the mesh accuracy in Lumerical was also increased from ``2'' to ``3'' for the plots shown in Fig. \ref{c2c_mode_linear}. The performance of the coupler as a function of wavelength and polarization can be found later in Fig. \ref{c2c_wvl_pol_low_index}. The data shows a 1-dB bandwidth exceeding 360 nm (1280-1640 nm) with polarization dependent losses below 0.1 dB between the fundamental TE and TM mode. Similarly, the dependence of GRIN chip-to-chip coupling as a function of the underfill epoxy refractive index is shown in Fig. \ref{c2c_epoxy_low_index}. The plot demonstrates low coupling loss below 1 dB for epoxy refractive indices ranging from 1.3-1.5, enabling design flexibility when selecting the UV curable epoxy as well as operational robustness to changes in index incurred by thermal fluctuations imposed by dynamic environments. Finally, the affect of randomized refractive index fluctuations for each layer on the overall chip-to-chip coupling efficiency is shown in Fig. \ref{hist_index_percent_error} and \ref{combined_percent_error}. Specifically, Fig. \ref{hist_index_percent_error} shows the raw distributions of the number of 2D simulations where a given coupling loss was calculated, and how these distributions change as a function of the maximum applied percentage error. This data was then re-analyzed to plot the change in the average coupling loss ($\mu$) with the maximum applied percentage error, and this is what Fig. \ref{combined_percent_error} is highlighting. The same process was followed for random fluctuations in layer thickness, and the results from this are also plotted in Fig. \ref{combined_percent_error}. The shaded regions represent one standard deviation about the mean, or $\pm \ \sigma$.


\subsection{Evanescent inter-layer coupler add-on for GRIN coupler}
\label{sec:ev_results}
\par
\begin{figure}[!ht]
\centering
\subfigure[\label{evanescent_schematic}]{\includegraphics[scale=0.19]{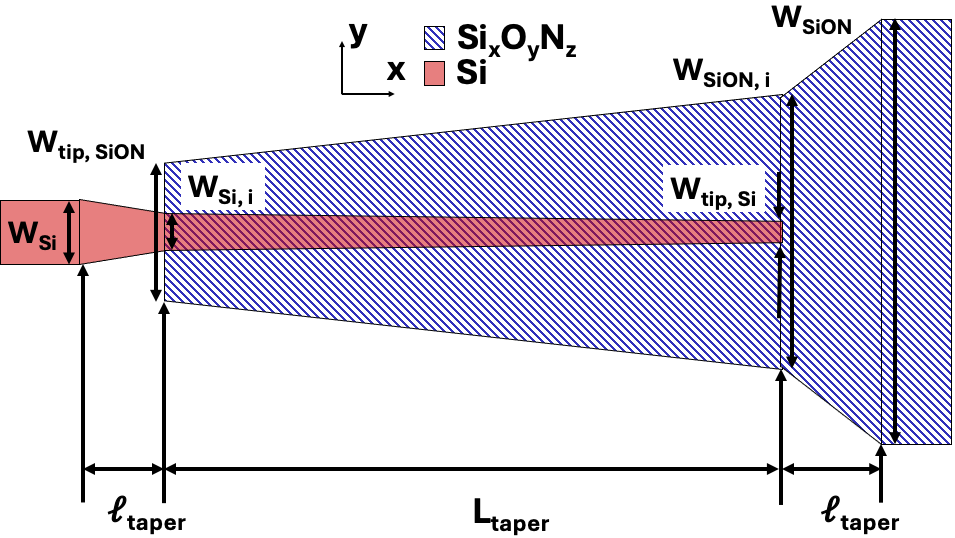}}
\subfigure[\label{evanescent_coupling_length}]{\includegraphics[scale=0.42]{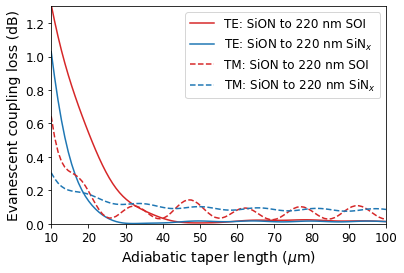}}
\subfigure[\label{evanescent_sion_to_si_mode_with_callouts}]{\includegraphics[scale=0.34]{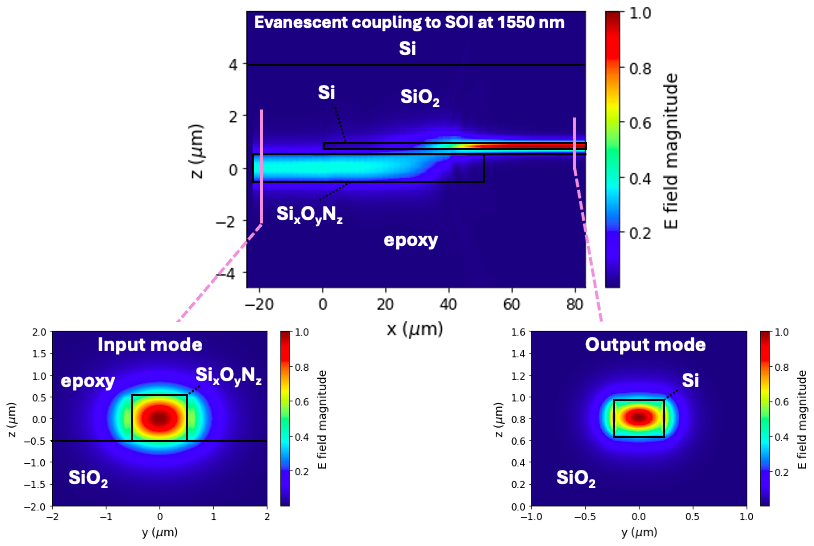}}
\caption{Simulations showing the performance of the evanescent coupler. In \subref{evanescent_schematic} a top down view shows the double taper design. In this figure, the label ``Si" (e.g. W$_{\text{Si,i}}$) can indicate either a SOI or SiN$_x$ waveguide. In \subref{evanescent_coupling_length} the loss through the coupler as a function of taper length ($L$) is shown. In \subref{evanescent_sion_to_si_mode_with_callouts} a side view from a 3D FDTD simulation shows the optical mode's transition from the SiON to an SOI waveguide.}
\label{evanescent_coupler}
\end{figure}

\begin{figure}[!t]
\centering
\subfigure[\label{c2c_wvl_pol_low_index}]{\includegraphics[scale=0.46]{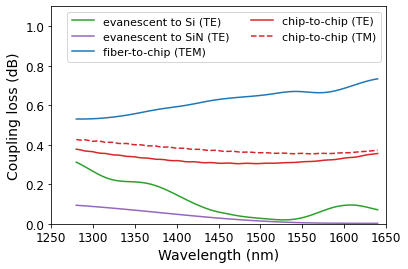}}
\subfigure[\label{c2c_epoxy_low_index}]{\includegraphics[scale=0.46]{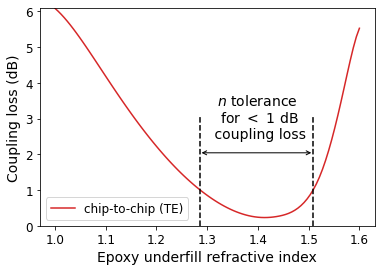}}
\subfigure[\label{hist_index_percent_error}]{\includegraphics[scale=0.41]{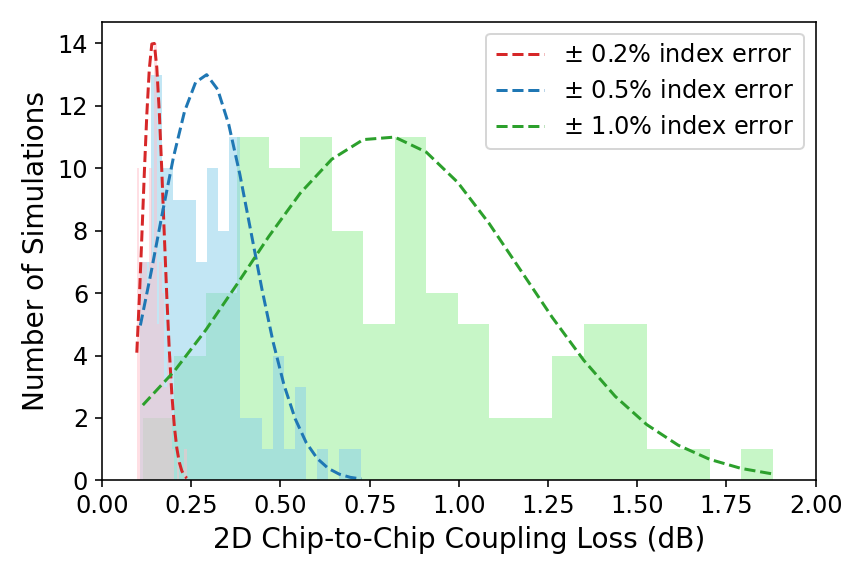}}
\subfigure[\label{combined_percent_error}]{\includegraphics[scale=0.46]{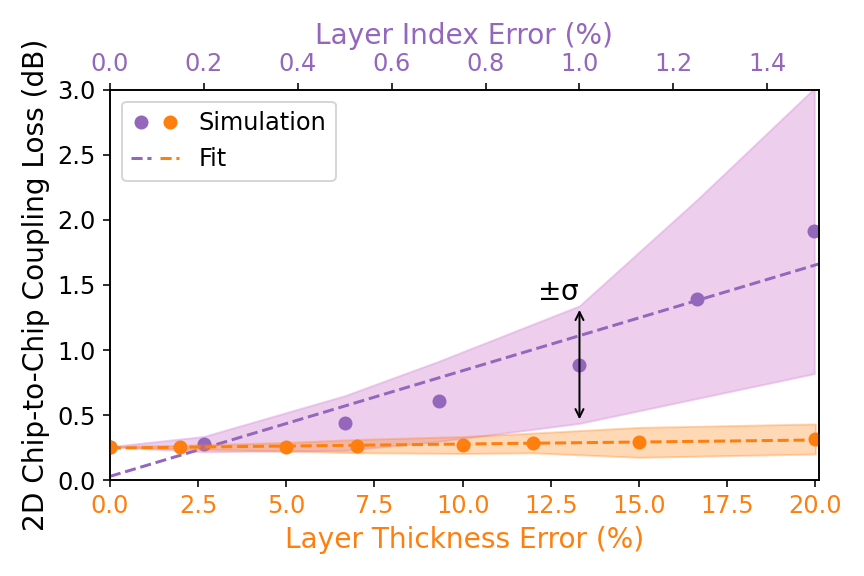}}
\caption{Simulations showing the effect of underfill refractive index, wavelength, and refractive index variation on GRIN coupling. In \subref{c2c_wvl_pol_low_index} the coupling loss is shown for the GRIN fiber-to-chip, GRIN chip-to-chip, and evanescent couplers as a function of wavelength. Note that the evanescent SOI tapers were of length $L$ = 40 $\mu$m and $\ell$ = 10 $\mu$m and the SiN$_x$ tapers were of length $L$ = 25 $\mu$m and $\ell$ = 10 $\mu$m. In \subref{c2c_epoxy_low_index} the chip-to-chip coupling loss is shown as a function of the underfill refractive index, showing a wide 1-dB index tolerance of approximately 0.2. In \subref{hist_index_percent_error}, histogram distributions show the number of 2D simulations as a function of the coupling loss. To obain these distributions, randomized refractive index errors were applied to the GRIN coupler (with each layer having a different random error applied to it) where the maximum possible percentage error for a given layer ranged from 0.2-1.0\%. The mean and standard deviation of each distribution was then plotted in \subref{combined_percent_error} as a function of the maximum percentage error and fit using linear regression.}
\label{c2c_wvl_performance}
\end{figure}

 The sweeps over W$_{\text{tip, SiON}}$, W$_{\text{tip, Si}}$, W$_{\text{SiON, i}}$, and W$_{\text{Si, i}}$ resulted in optimized values of 400 nm, 100 nm, 1200 nm, and 230 nm, respectively, for coupling from SiON into SOI at 1550 nm with a 200 nm taper-to-taper gap. For coupling from SiON into SiN$_x$ with the same simulation parameters, the optimized values for W$_{\text{tip, SiON}}$, W$_{\text{tip, SiN}}$, W$_{\text{SiON, i}}$, and W$_{\text{SiN, i}}$ were 400 nm, 200 nm, 1200 nm, and 1000 nm. A plot showing the results for evanescent coupling loss versus $L_{\text{taper}}$ for the optimized design can be found in Fig. \ref{evanescent_coupling_length}. The results show that to achieve coupling loss below 0.05 dB ($>$ 99\% efficiency), the adiabatic taper length should be $>$ 25 $\mu$m and $>$ 40 $\mu$m for transfer from SiON into SiN$_x$ or SOI waveguides, respectively. The transformation of the TE mode at 1550 nm from the SiON waveguide to a SOI wavguide can be seen in Fig. \ref{evanescent_sion_to_si_mode_with_callouts} using 3D FDTD simulations. The plot demonstrates high efficiency coupling using the double taper layout without substrate leakage through the BOX layer. Lastly, a plot showing the coupling loss as a function of wavelength can be found in Fig. \ref{c2c_wvl_pol_low_index}. The data represents a 1-dB bandwidth of more than 360 nm. 

\section{Discussion of GRIN coupler dependencies}
\label{sec:discussion}

\par The primary factors contributing to the GRIN coupling loss are scattering losses associated with the GRIN facet interface and the output waveguide interface. Both of these losses can be attributed to imperfect modal overlap due to the parabolic GRIN profile. In the case of fiber-to-chip coupling, the GRIN facet interface involves modal overlap between the SMF and the GRIN coupler. Here, the shape of the mode inside the GRIN lens is not perfectly circular, like the mode inside the fiber as seen in Fig. \ref{f2c_mode_linear}, due to the GRIN profile having a high index material at the bottom ($n_H$ = 1.7) and being rectangular in shape. In the case of chip-to-chip coupling, the modal overlap at this interface is increased since the structures are identical to one another (i.e. symmetric). The increased modal overlap at the GRIN facet interface accounts for the improvement in maximum coupling efficiency seen by the chip-to-chip coupler versus the fiber-to-chip coupler.

\par Similarly, for both fiber-to-chip and chip-to-chip coupling, the output waveguide interface involves modal overlap between the GRIN lens output and the output SiON waveguide. These modes are not perfectly matched, as depicted in Fig. \ref{f2c_mode_linear} and \ref{c2c_mode_linear}. This is due to the fact that the output waveguide layer and the bottom layer of the GRIN lens are the same. A GRIN coupler which can separate the GRIN lens profile (including refractive indices and layer thicknesses) from the output waveguide profile, specifically by having a higher index for the bottom layer (and thus a higher vertical confinement), may result in a higher coupling efficiency as described in \cite{2021weninger}. However, obtaining a different refractive index and thickness for the output waveguide and the bottom layer of the GRIN lens increases the fabrication difficulty significantly, so it was not simulated in this study.

\subsection{Effect of lens radius and GRIN length}
\par The wide design and fabrication tolerance with respect to $r$ and $L$ as depicted by Fig. \ref{f2c_sweep_low_index} and \ref{c2c_sweep_low_index} can be explained by the the total GRIN focal length remaining approximately equal within the 1-dB coupling loss contour. As the lens radius increases, $\ell$ decreases, and thus the length of the tapered section of the GRIN, $L$, must increase to achieve the same total GRIN length equal to $\ell$ + $L$. As the lens radius continues to increase, the GRIN facet becomes increasingly flat. When the GRIN coupler is flat, the horizontal confinement becomes entirely due to the tapered nature of the GRIN, whose length is determined by the focal length. Because the GRIN taper is only 20-30 $\mu$m long and thus non-adiabatic, the coupling efficiency decreases, as described by the data for a 100 $\mu$m lens radius in Fig. \ref{f2c_sweep_low_index} and \ref{c2c_sweep_low_index}. In other words, the use of a lens allows for independent control of the horizontal and vertical confinement which results in a higher coupling efficiency.

\subsection{Effect of wavelength dependence on GRIN coupler performance}
\par Another aspect to explain is the wavelength response of the fiber-to-chip coupler, which showed a slight decrease in coupling efficiency from 1280 nm to 1640 nm based on Figure \ref{c2c_wvl_pol_low_index}. The improvement with decreasing wavelength can be attributed to the fiber MFD more closely matching the GRIN MFD for shorter wavelengths. The weak wavelength dependence highlights how the focal length of the GRIN lens is independent of wavelength. This is because the focal length depends on the ratio between $n_H$ and $n_L$ and the thickness $T$, as described in Section \ref{sec:methods} and by equation \ref{equation2}, which remain approximately constant across the infrared spectrum for SiON films. An identical argument can be made for the chip-to-chip coupler, which sees a nearly negligible wavelength dependency. Finally, the SOI and SiN$_x$ evanescent couplers also demonstrate a relatively low wavelength dependence from 1280-1640 nm, with both couplers showing a higher coupling loss for shorter wavelengths. This can be explained by the fact that for shorter wavelengths, the evanescent field through the tapers decreases, resulting in a lower interaction. Along the same lines, the SOI-SiON system shows a higher sensitivity than the SiN$_x$-SiON system due to the SOI-SiON modal effective refractive index having a stronger wavelength dependence compared to the SiN$_x$-SiON mode, leading to a decrease in the evanescent field strength and an increase in coupling loss.

\subsection{Effect of epoxy and layer refractive index variation}
\par While the GRIN coupler demonstrated a wide 1-dB tolerance to epoxy refractive index changes of approximately $\pm$ 0.1 from the initial value of $\sim$ 1.44, there is a clear increase in coupling loss for refractive indices $<$ 1.3 or $>$ 1.5. For epoxy refractive indices $>$ 1.5, the epoxy begins to become a higher index than $n_L$, the lowest refractive index layers of the GRIN. For layers which have a lower refractive index than the epoxy refractive index, they effectively become part of the cladding and no longer contribute to GRIN lensing, causing n$_L$/n$_H$ to approach unity and $\alpha \rightarrow$ 0 according the equation \ref{equation2}. The GRIN focal length effectively increases and coupling loss increases as a result. An analogous argument can be made for indices $<$ 1.3 - n$_L$/n$_H$ effectively increases causing $\alpha$ to increase (and the GRIN period to decrease) causing additional coupling losses.

\par The introduction of random fluctuations on the layer refractive index also induced additional chip-to-chip coupling losses as evidenced by Figure \ref{hist_index_percent_error} and \ref{combined_percent_error}. The coupler was more sensitive to index fluctuations compared to thickness fluctuations since the GRIN lens period is directly correlated to the refractive index profile (it is proportional to $\alpha^{-1}$) and correlated indirectly with layer thickness (t$_{\text{layer}}$) through the total thickness $T$ ($\alpha \propto T^{-2}$). The 1-dB tolerance for refractive index error per layer was $\sim$ $\pm$ 1\% based on Figure \ref{combined_percent_error}. This equates to approximately $\pm$ 0.015-0.017 in terms of the refractive index control necessary during the PECVD process. While $\pm$ 1\% seems like extremely tight control, data for similar processes such as inductively coupled plasma (ICP) PVD has standard refractive index control of $<$ $\pm$ 0.3\% over $>$ 75 wafers using commercially available tools \cite{2023oxford}. On the other hand, the 2D chip-to-chip coupling loss was essentially unchanged by layer thickness errors as large as $\pm$ 20\%, or roughly $\pm$ 200 nm.

\subsection{Effect of refractive index on evanescent coupler taper length}
\par The evanescent coupler design used in this study was described extensively in \cite{2023weninger, 2024weninger} for SOI to SiN$_x$ coupling. However, one point to elaborate on in is the improvement seen in SiN$_x$ compared to SOI for the same adiabatic taper length as depicted by Fig. \ref{evanescent_coupling_length}. This coupling efficiency improvement can be attributed to the lower refractive index of SiN$_x$ compared to the SOI, which more closely matches that of the SiON. The lower refractive index means larger feature sizes can be used to achieve the same effective refractive index in SiN$_x$. Thus, these larger lateral feature sizes mean the shape of the mode in the SiN$_x$ more closely matches that of the SiON prior to the mode entering the adiabatic region of the taper. Therefore, a faster transition can be used for SiON to SiN$_x$ compared to SiON to SOI to achieve the same coupling loss. 

\begin{table}[!t]
\captionsetup{font=normalsize}
\centering
\renewcommand{\arraystretch}{1.3}
\caption{Summary of loss contributions through GRIN fiber-to-chip, chip-to-chip couplerm and evanescent coupler.} 
\small
\begin{tabular*}{0.4\textwidth}{|c|c|c|}
\cline{1-3}
\textbf{Component} & \multicolumn{2}{|c|}{\textbf{Loss (dB)}}\\
\cline{1-3}
Final waveguide material & SOI & SiN$_x$\\
\cline{1-3}
\multicolumn{3}{|c|}{Fiber-to-chip coupler}\\
\cline{1-3}
Single GRIN lens & \multicolumn{2}{|c|}{0.62}\\

\cline{1-3}
Evanescent coupler$^\dagger$ & 0.03 & 0.04 \\
\cline{1-3}
\textbf{Total} & \textbf{0.65} & \textbf{0.66} \\
\cline{1-3}
\multicolumn{3}{|c|}{Chip-to-chip coupler}\\
\cline{1-3}
Two GRIN lenses & \multicolumn{2}{|c|}{0.27}\\
\cline{1-3}
Evanescent couplers$^{\dagger*}$ & 0.06 & 0.08 \\
\cline{1-3}
\textbf{Total} & \textbf{0.33} & \textbf{0.35} \\
\cline{1-3}
\end{tabular*}
\vskip1em
\begin{flushleft}
$^\dagger$Assuming a 40 $\mu$m and 25 $\mu$m long adiabatic taper for SOI and SiN$_x$, respectively\\
$^*$Includes evanescent coupling loss at GRIN lens input and output
\end{flushleft}
\label{tab:total_loss}
\end{table}
\section{Comparison to other types of chip-to-chip couplers}
\begin{figure}[!t]
\centering
\subfigure[\label{c2c_star_plot}]{\includegraphics[scale=0.52]{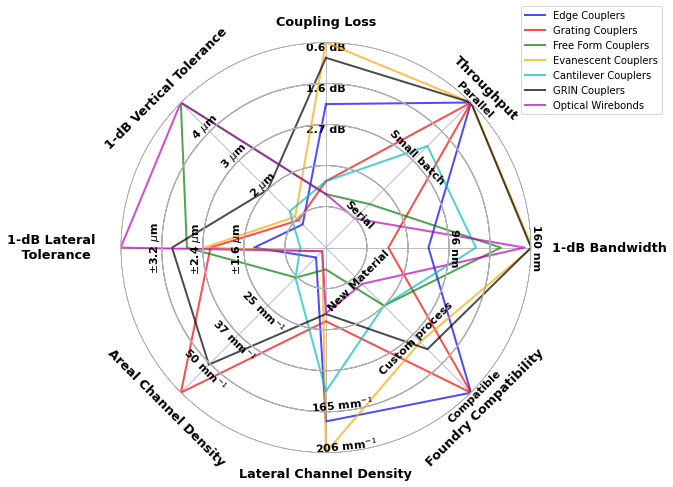}}
\caption{A radar plot comparing different chip-to-chip coupling strategies in silicon photonics in terms of performance metrics relevant to data- and tele-communications applications and scaling to high volume manufacturing.}
\label{star_plot}
\end{figure}

\par The total coupling loss was computed for fiber-to-chip coupling from an SMF to SOI or SiN$_x$ and for chip-to-chip coupling from SOI to SOI or SiN$_x$ to SiN$_x$ (or any combination thereof), can be found in Table \ref{tab:total_loss} for 1550 nm wavelength. From the data in the table, it can be see that the fiber-to-chip coupling loss of 0.65 dB is $<$ 1 dB, the chip-to-chip coupling loss of 0.33 dB is $<$ 1 dB, and their combination of 0.98 dB is also $<$ 1 dB. An important implication of this data is that the GRIN coupler design can support optical fanout to Pbps optical I/O with a sub-dB aggregate interfacial loss contribution to the total loss budget using the layout described by Fig. \ref{fig:overview}. A brief review of the performance metrics for alternate vertical couplers are presented in Table \ref{tab:review}. In comparing the values for coupling loss, alignment tolerance, and bandwidth associated with the GRIN structure from this study, our coupler stacks up well to peer innovations. First, in terms of coupling loss, our GRIN chip-to-chip coupler displayed a maximum coupling efficiency of slightly greater than 93\%, or 0.27 dB of loss. This is competitive compared to the other high performance couplers which also have coupling losses below 1 dB. Second, our widened translational alignment tolerances above $\pm$ 2.3 $\mu$m compare favorably to other evanescent, edge, or grating coupler designs which are typically less than 2 $\mu$m. A lateral alignment tolerance of $\pm$ 2.3 $\mu$m is wide enough to use automated pick-and-place die bonders for passive assembly in high volume manufacturing \cite{mrsi_s_hvm}. When it comes to footprint, the $<$ 80 $\mu$m length and 11 $\mu$m width of this design makes it compact, especially compared to evanescent designs requiring greater than 1 mm in length to achieve a lower than $\pm$ 2 $\mu$m lateral tolerance. 

\par One additional important point to keep in mind when comparing our alignment tolerances to a free form coupler, which can achieve exceptionally large vertical alignment tolerances, is the monolithic, planar nature of our design which relies on standard, controlled microelectronics foundry processes and materials instituted in the BEOL. This is contrasted by couplers which are fabricated using TPP techniques, a foundry incompatible process with significantly lower throughput due to the sequential patterning of optical components with repeated exposures instead of processing devices in parallel. In terms of material properties, it also means our coupler is reflow temperature compatible and not subject to the same mechanical, thermal, or humidity based reliability problems which may arise for polymer based structures. Additionally, because the GRIN coupler width was 11 $\mu$m and the optical mode was highly confined within the GRIN, even during expansion, our lateral pitch can be made significantly compared to free form couplers which expand the mode to $>$ 100 $\mu$m in diameter. A finer I/O pitch directly equates to a higher I/O density along the shoreline of the Si PIC, and thus a higher possible data capacity. Thus, our coupler may provide more rapid, less costly scaling to high volume manufacturing applications requiring high density integration which is especially crucial in the context of scaling to greater than 1 Pbps data- or tele-communications co-packaged systems. 

\definecolor{LightBlue}{rgb}{0.78, 0.95, 1}
\begin{table}[!t]
\footnotesize
\centering
\caption{Summary of recent examples of chip-to-chip optical coupling schemes. This list does not include all chip-to-chip coupling schemes and is only meant to yield a general idea of competing designs and performance.} 
\renewcommand{\arraystretch}{1.2}
\newcolumntype{C}[1]{>{\centering\arraybackslash}m{#1}}
\scriptsize
\begin{tabular*}{\textwidth}{C{1.55cm}|C{1.1cm}|C{1.75cm}|C{1.2cm}|C{0.9cm}|C{1.9cm}|C{0.9cm}|C{0.6cm}}
\hline
\multirow{2}{1.55cm}{\centering \textbf{Package}} & \multirow{2}{1cm}{\centering \textbf{Coupler}} & \multirow{2}{1.2cm}{\centering \textbf{CL$^1_{1550}$(dB)}} & \multicolumn{2}{c|}{\textbf{1-dB Tolerance$^2$ ($\mu$m)}} & \multirow{2}{1.9cm}{\centering \textbf{1-dB BW$^3$ (nm)}} & \multirow{2}{0.9cm}{\centering \textbf{LxW($\mu$m)}} & \multirow{2}{0.6cm}{\centering \textbf{Ref.}}\\
\cline{4-5}
& & &\textbf{Lateral} & \textbf{Vertical} & & &\\
\hline
InP to SOI & Edge & 0.25 & \cellcolor{LightBlue} $\pm$ 0.7 & \cellcolor{LightBlue} $\pm$ 0.5  &  120 (1510-1630) & 75x0.4 & \cite{2017barwicz} \\
\hline
InP to SOI & Edge & 0.9 & \cellcolor{LightBlue} $\pm$ 1.7 & \cellcolor{LightBlue} $\pm$ 0.5 &  120 (1510-1630) & 250x0.4 & \cite{2017barwicz} \\
\hline
SOI to SOI & PWB & 1.6 &/&/& $>$ 300 (1280-1580)  & 2 (W) & \cite{2012lindenmann}\\
\hline
SOI to polymer &Evanescent& 1.25/0.5 (TE/TM)\newline $<$ 0.5 (TE/TM)  & $\pm$ 2 \newline / & / & \makecell[b]{$\pm$ 60 (1310) \\ $\pm$ 70 (1550)}  & 1500x6.5 &\cite{2018dangel}\\
\hline
\rowcolor{LightBlue}
Polymer to polymer & Free form &  \makecell[b]{0.22 (TE) \\0.25 (TM)}  & $\pm$ 1.3 &  35 & $\pm$ 200 (850) & 10x30 &\cite{2020shaoliang}\\
\hline
\rowcolor{LightBlue}
SOI to polymer& Evanescent&0.2 & $>$ $\pm$ 5 \newline $\pm$ 1.5\textdegree(yaw) & 0.5  & $\pm$ 100 (1550) & 200x15.3 & \cite{2021englund}\\
\hline
\rowcolor{LightBlue}
SOI to IOX & Evanescent &$<$ 1 (TE/TM) & $\pm$ 4 & 3 & $\pm$ 30 (1550) & 1500x12 & \cite{2021brusberg}\\
\hline
Si$_3$N$_4$ to IOX & Evanescent &0.7 & $>$ $\pm$ 4 & 
$<$ 2 & 75 (1515-1590) & 2000x11 &\cite{2023brusberg}\\
\hline
Si$_3$N$_4$ to Si$_3$N$_4$ &Evanescent &0.54 & $<$ $\pm$ 2 & $<$ 0.8 & 400 (1200-1600) & 1000x3 & \cite{2024zhang}\\
\hline
SOI to SOI & Grating & 0.94 & / & / & 21 (1539-1560) & 20x20 & \cite{2024wang} \\
\hline
Si$_3$N$_4$ to SOI & Evanescent & 0.39 (TE) & $\pm$ 1.56 & $>$ 1.1& 160 (1480-1640) & 520x1 &\cite{2024weninger}\\
\hline
\rowcolor{LightBlue}
\textbf{SiN$_x$ to SiN$_x$} & \textbf{GRIN} & \makecell[b]{0.33 (TE) \\ 0.41 (TM)} & \textbf{$\pm$ 2.38} &\textbf{$\pm$ 2.24 \newline 11 (rise)} & $>$ 360 (1280-1640) & \textbf{66x11} &\textbf{this work}\\
\hline
\rowcolor{LightBlue}
\textbf{SOI to SOI} & \textbf{GRIN} &\makecell[b]{0.35 (TE) \\ 0.43 (TM)} & \textbf{$\pm$ 2.38} &\textbf{$\pm$ 2.24 \newline 11 (rise)} & $>$ 360 (1280-1640) & \textbf{51x11} &\textbf{this work}\\
\hline
\end{tabular*}
\vskip1em
\begin{flushleft}
$^1$CL$_{1550}$ = coupling loss at 1550 nm, $^2$Tolerance = misalignment tolerance, $^3$BW = bandwidth
\end{flushleft}
\label{tab:review}
\end{table}

\section{GRIN chip-to-chip coupler fabrication progress}\label{sec:fab}
\par
\begin{figure}[!t]
\centering
\includegraphics[scale=0.35]{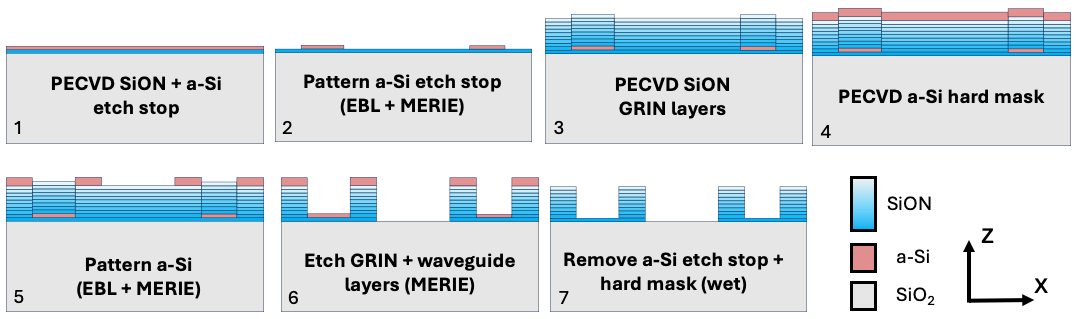}
\caption{The process flow for fabrication of thick GRIN structures on glass substrates using EBL. This figure only shows the process steps which were executed in this report. The additional process steps which would follow step 7 would include deep dry etching to form glass edge facets for edge coupling and the flip-chip pick-and-place sequence using UV curable epoxy. These steps have been developed for flip-chip evanescent coupling as shown in \cite{2024weninger}, and future work will include transferring these processes to the GRIN prototype.}
\label{grin_process_flow}
\end{figure}

 Prior work has demonstrated GRIN fiber-to-chip couplers with integrated lenses having a coupling loss of 0.4 dB from 1530-1625 nm \cite{2007sun}. This work intends to add to the prior work by increasing the thickness of the GRIN lens from 6 $\mu$m to 11 $\mu$m in order to couple light from flat-faceted, SMF28 fibers. In addition, the fabrication of GRIN lenses on glass substrates was explored as this is useful to the development of high performance interposers and package substrates with integrated silicon based waveguides. These two additions to prior work - an increased GRIN thickness and the use of glass substrates - presented fabrication challenges which needed to be overcome. First, a thicker GRIN lens necessitated patterning and etching a thicker aSi hard mask without losing resolution. Second, due to available tool constraints, electron beam lithography (EBL) processes for thick photoresist patterning on insulating substrates needed to be developed. This work builds on the prior work in \cite{2024weninger} which established processes for the patterning of thin low pressure chemical vapor deposition (LPCVD) SiN$_x$ waveguides on glass substrates. The targeted process flow is shown in Fig. \ref{grin_process_flow} for creating the thick fiber-to-chip and chip-to-chip GRIN coupling structures on glass. This process flow can be duplicated for silicon substrates as well, assuming they contain the standard 2-3 BOX layer. It is important to note that, while the processes shown in Fig. \ref{grin_process_flow} represent the fabrication work done in this study, additional process steps will be necessary to complete a working GRIN prototype. The most relevant of these additional steps are the deep dry etching of the glass edge facets and passive assembly using flip-chip pick-and-place die bonding with UV curable epoxy. These processes were established for flip-chip evanescent coupling and for edge coupling to LPCVD SiN$_x$ standard tapers on glass in \cite{2024weninger}, and the application of these working processes to the GRIN coupler will be saved for future work. Thus, the following sections describe process development efforts intended to establish fabrication steps that match the performance of prior work and advance to thicker GRIN lenses on glass substrates.

\subsection{PECVD of SiON films with controllable refractive index}
\par The PECVD processes for SiON films with a controllable refractive index were developed using the a Samco PD220NL tool. The recipe used for deposition included the following parameters: 5 minutes, 0-300 sccm N$_2$O, 10 sccm SiH$_4$, 2 sccm NH$_3$, 390 sccm N$_2$, 13.56 MHz frequency, 350\textdegree C, 80 Pa, and 100 W RF bias. The Samco PD220NL tool cannot reach high enough plasma densities to dissociate N$_2$; therefore, NH$_3$ needed to be added as a precursor. Depositions were carried out at 350\textdegree C on a 6" AlN chuck with Si pieces approximately 15 mm by 15 mm in size which were cleaved from a 100 mm Si wafer ($<$100$>$, P type, 500 $\mu$m thick). Each deposition was carried out for the same time (5 minutes), but with a different N$_2$O flow rate. The refractive index and thickness of the films were measured using a Semilab SE-2000 spectroscopic ellipsometer. Each sample was then cross sectioned and the thickness of the thin film was measured using a scanning electron microscrope (SEM). When fitting the refractive index, the thickness of the films was checked to match against the SEM measurements in order to provide a more accurate determination of the refractive index. Determining the refractive index and thickness of each film was done by measuring the reflected polarization spectrum at three different angles for wavelengths from 400 nm to 1600 nm, and modeling the film using the two lowest order Cauchy coefficients (i.e. $n(\lambda) = n_{\infty} + n_1(\lambda)$). The resultant data for refractive index and thickness as a function of N$_2$O flow rate can be found in Fig. \ref{GRIN_experimental_index_plot} and \ref{GRIN_experimental_thickness_plot}. The work in this section demonstrated successful fabrication of refractive index controllable SiON films using PECVD with standard, foundry compatible precursors and metrology techniques.

\subsection{Electron beam lithography of thin SiON film using aSi mask}
\label{sec:etch_stop}
\par
\begin{figure}[!t]
\centering
\subfigure[\label{refractive_index_heatmap}]{\includegraphics[scale=0.45]{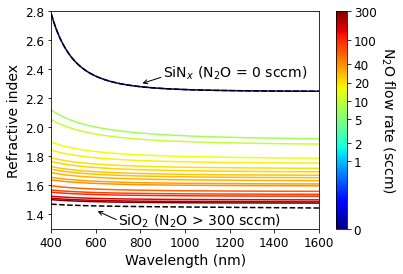}}
\subfigure[\label{GRIN_experimental_thickness_plot}]{\includegraphics[scale=0.45]{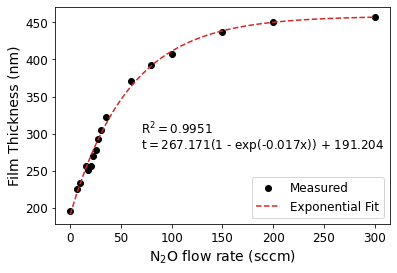}}
\subfigure[\label{GRIN_experimental_index_plot}]{\includegraphics[scale=0.45]{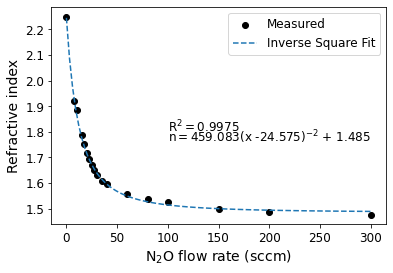}}
\subfigure[\label{grin_etch_stop_SEM}]{\includegraphics[scale=0.37]{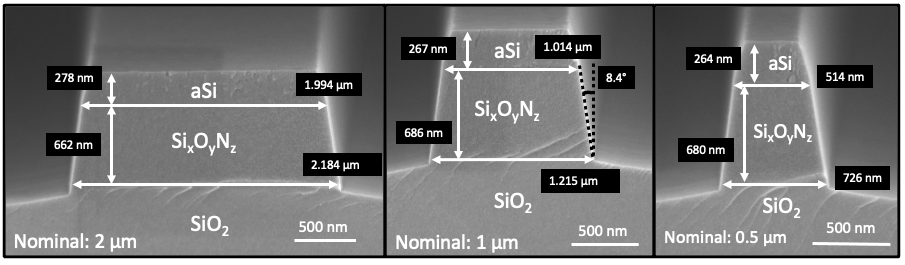}}
\caption{Experimental data for fabrication of the GRIN coupler. In \subref{refractive_index_heatmap} a plot of the measured refractive index of the SiON films as a function of wavelength for different N$_2$O gas flow rates. In \subref{GRIN_experimental_thickness_plot} and \subref{GRIN_experimental_index_plot} the thickness and refractive index, respectively, of the SiON film as a function of N$_2$O flow rate are shown. In \subref{grin_etch_stop_SEM}, cross sectional SEM images show the waveguide profile for different waveguide widths prior to removal of the aSi etch stop.}
\label{grin_index_thickness_layer1}
\end{figure}

 Following establishment of the refractive index profile, the processes for patterning the amorphous Si (aSi) etch stop on top of the waveguide layer were developed. For this process, 700 nm SiON was deposited using a N$_2$O flow rate of 10.9 sccm ($n = 1.898$ at 1550 nm) on a 3 $\mu$m SiO$_2$ BOX, following by the deposition of 400 nm of aSi using the same tool. The recipe for aSi deposition: 8 min 20s, 28 sccm SiH$_4$, 500 sccm Ar, 100 W RF, 80 Pa, 270\textdegree C, and 13.56 MHz frequency. The EBL to pattern the aSi builds off of the protocol established in \cite{2024weninger} for the EBL of thin SOI and SiN$_x$-on-glass films using a diluted AZ nLOF 2020 photoresist. The primary change that was made was to the photoresist dilution, which was changed to 3:2 (nLOF:PGMEA) in order to increase the resist thickness to approximately 800 nm for a 2000 rpm, 60s spin. This change was made because a 400 nm PECVD aSi hard mask was used compared to prior studies which used a 250 nm LPCVD poly-Si hard mask. The thicker etch stop was used due to the lower etch resistance of the PECVD aSi compared to LPCVD poly-Si. The aSi was etched using an Applied Materials Precision 5000 RF generated, magnetically enhanced RIE (MERIE) tool with a Cl$_2$/HBr gas chemistry. The Cl$_2$ recipe was as follows: 60 sccm Cl$_2$, 20 sccm HBr, 100 mTorr, 250W, 60 Gauss.  The selectivity of the photoresist was determined to be 1.58 (aSi:PR). The SiON was etched using a different chamber of the same tool with CF$_4$/CHF$_3$/Ar chemistry. The CHF$_3$ recipe was as follows: 155 seconds, 55 sccm CHF$_3$, 45 sccm CF$_4$, 100 sccm Ar, 50 mTorr, 600W, 60 Gauss. The selectivity of the aSi was determined to be 9.4 (aSi:SiON). Cross sectional SEM images showing the SiON waveguide with the aSi hard mask still remaining can be found in Fig. \ref{grin_etch_stop_SEM}, achieving widths from 500 nm to $>$ 1 $\mu$m with a measured sidewall slope of 8.4\textdegree. The optimized dose for this structure was 55 $\mu$C/cm$^2$ with a beam current of 10 nA using an Elionix HS-50, 50 keV accelerating voltage system. The work in this section demonstrated successful EBL patterning of SiON waveguides using a PECVD aSi hard mask and a diluted, negative tone chemically amplified photoresist.

\subsection{Electron beam lithography of thick SiON film on glass using aSi mask}

\par With the refractive index profile and patterning of the aSi etch stop determined, the next step was to develop a process for patterning the GRIN lens that was $>$ 10 $\mu$m thick. A similar process to \cite{2007sun} was used, with the exception of EBL in place of i-line lithography. This was not done due to small features - the smallest feature size of the GRIN lens design is $>$ 1 $\mu$m - but rather the lack of access to a repeatable process for patterning 1 $\mu$m features using the UV photolithography tools within the facilities used. Furthermore, it was known based on prior work that alignment with $<$ 0.5 $\mu$m precision (and preferably $<$ 0.2 $\mu$m precision) between the GRIN lens pattern and the etch stop pattern would be critical to achieving the low coupling loss predicted by simulation.

\par Like Section \ref{sec:etch_stop}, we followed a process similar to that used in \cite{2024weninger} for EBL using AZ nLOF 2000 series photoresist. It was anticipated based on the results in Section \ref{sec:etch_stop} that for Cl$_2$/HBr etching of aSi, the selectivity was approximately 1.58 (aSi:PR). Similarly, based on literature \cite{1996bazylenko} and prior test samples, it was determined that faceting was occurring in the corners of the aSi hard mask as a result of etching the presence of a CHF$_3$ and Ar gas chemistry. Due to the low selectivity and presence of faceting, an aSi hard mask of 4 $\mu$m and a photoresist mask of 4 $\mu$m were targeted. The 4 $\mu$m aSi hard mask was deposited using the same PECVD recipe as in Section \ref{sec:etch_stop}, but for 80 minutes. The 4 $\mu$m thick photoresist was achieved using the same EBL steps as in Section \ref{sec:etch_stop}, but using undiluted AZ nLOF 2035 with a 2500 rpm, 30s spin recipe. The development time was also adjusted to be 4 min. The SiO$_2$ pieces were then mounted on a 150 mm Si carrier wafer as in Section \ref{sec:etch_stop} for subsequent etching. The Cl$_2$/HBr etch for the aSi hard mask was identical to that of Section \ref{sec:etch_stop} aside from the time. The fluorine based etch was altered to a CHF$_3$/Ar only gas chemistry to increase selectivity to the aSi hard mask for the thicker GRIN lens. Based on prior results \cite{1996bazylenko, 2007sun}, this change was anticipated to have little effect on the etch profile of the waveguide layer. The etch recipe that was used for the GRIN lens was as follows: 91 minutes, 55 sccm CHF$_3$, 100 sccm Ar, 50 mTorr, 600W, 60 Gauss. The average selectivity of the aSi was determined to be 5.86 (aSi:SiON). 

\par
\begin{figure}[!ht]
\centering
\subfigure[\label{grin_eds_with_SEM}]{\includegraphics[scale=0.27]{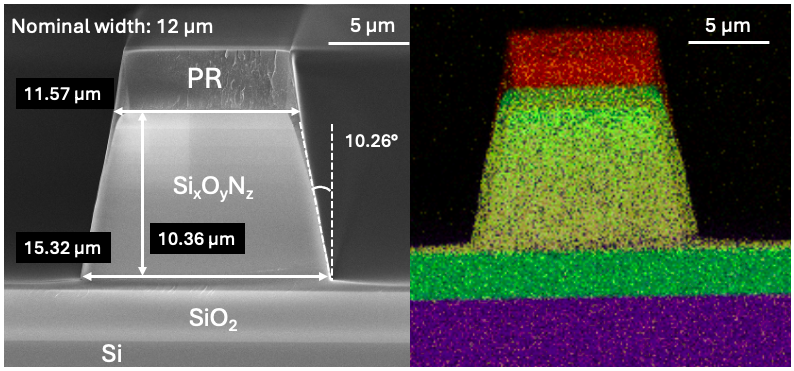}}
\subfigure[\label{EDS_plot}]{\includegraphics[scale=0.39]{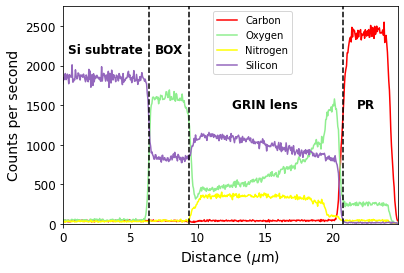}}
\subfigure[\label{hard_mask_on_glass}]{\includegraphics[scale=0.41]{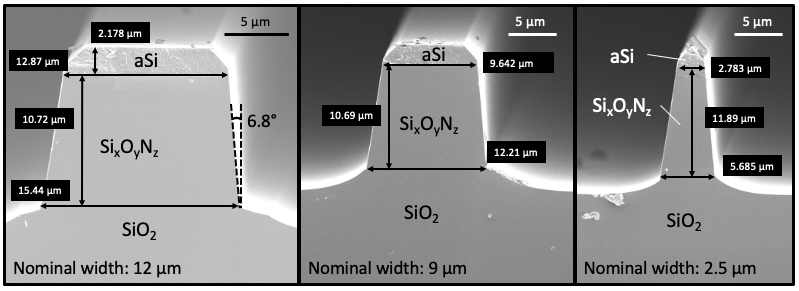}}
\subfigure[\label{hard_mask_on_glass_correct_size}]{\includegraphics[scale=0.41]{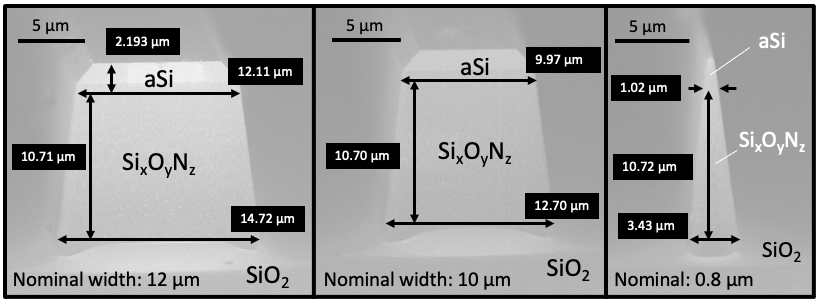}}
\caption{Experimental data showing fabrication of the GRIN lens. In \subref{grin_eds_with_SEM} a cross sectional SEM image alongside a 2D EDS map are shown for a GRIN lens test sample which used a photoresist mask only. In \subref{hard_mask_on_glass} cross sectional SEM imagse for a GRIN lens test sample which successfully EBL to pattern the aSi hard mask and GRIN stack on a fused silica substrate. In \subref{hard_mask_on_glass_correct_size} SEM images show the GRIN lens with the correct dimensions down to 1 $\mu$m following CHF$_3$ etching.}
\label{grin_cross_section}
\end{figure}

 Several sets of SEM images can be found in Fig. \ref{grin_cross_section} which show different aspects of the GRIN lens experimentally realized. For example, from the cross sectioned test sample shown in Fig. \ref{grin_eds_with_SEM} (which used a purely photoresist mask but contained the same SiON stack), the elemental composition of the GRIN lens using energy dispersive spectrometry (EDS) can be seen. The EDS results, which also included a line scan taken from bottom to top and shown in Fig. \ref{EDS_plot}, demonstrate a parabolically increasing amount of elemental oxygen when going from the bottom to the top of the GRIN lens. This reflects the parabolic increase of N$_2$O flow rate that was used for depositing each subsequent layer, which directly relates to refractive index. On the other hand, Fig. \ref{hard_mask_on_glass} shows a cleaved cross section with the fused silica substrate fully visible and the successful EBL patterning of a $>$ 10 $\mu$m stack on glass clearly represented. In addition, Fig. \ref{hard_mask_on_glass} reveals the faceting nature of the aSi hard mask mentioned earlier. Despite the faceting, the 4 $\mu$m thick aSi hard mask survived the deep etch without lateral dimension erosion affecting the stack beneath it. This is evidenced by smaller features, such as the 2.5 $\mu$m line shown in Fig. \ref{hard_mask_on_glass}, maintaining shape. Finally, additional SEM images in Fig. \ref{hard_mask_on_glass_correct_size}, which were not full cross sections but still taken from the edge, show proper etch dimensions compared with nominal values. Specifically, the images show the top width of the GRIN lens ranging from 1 $\mu$m to 12 $\mu$m and aligning well with the design values, along with a measured sidewall slope of 6.8\textdegree \ for a 10.72 $\mu$m thick stack. Based on these results, which were part of a dose matrix, an optimal EBL recipe of 25 $\mu$C/cm$^2$ for a 10 nA beam current was determined. The work in this section represent a step forward in terms of GRIN coupler fabrication as $>$ 10 $\mu$m thick lenses were able to be patterned on insulating substrates using EBL with a novel photoresist mixture.

\section{Conclusion}
In this study, the design and simulation of a novel GRIN coupler for chip-to-chip coupling was presented and demonstrated to achieve sub-dB coupling loss with broadband and alignment tolerant operation. This high performance can be achieved while coupling over a significant vertical chip-to-chip gap of $>$ 10 $\mu$m. In terms of adoption to high volume manufacturing, the coupler can be fabricated during, or after the completion of, typical BEOL steps using standard, low-temperature foundry compatible materials and processes, and can be used to connect waveguides standard to silicon photonic PDKs. The coupler can be fabricated at the wafer or panel level using parallel, standard thin film fabrication methods, and can be packaged using passive assembly with automated flip-chip die bonders. When placed at a 20 $\mu$m pitch, this coupler offers a density of 50 couplers/mm, a 6$\times$ improvement over systems directly connecting fibers to Si-PICs. In essence, this coupler will help enable the mass parallelization and cost-effective scaling of optical I/O in co-packaged optics systems. 

\begin{backmatter}

\bmsection{Acknowledgements}
\par This work was carried out in part through the use of MIT.nano's facilities and was supported by NSF Convergence Accelerator Track I: FUTUR-IC: A Sustainable Microchip Manufacturing Alliance, Award Number ITE-2345076.

\bmsection{Disclosures}
The authors declare no conflicts of interest.

\bmsection{Data availability} Data underlying the results presented in this paper are not publicly available at this time but may be obtained from the authors upon reasonable request.
\end{backmatter}

\bibliography{references}

\begin{thebibliography}{10}
\newcommand{\enquote}[1]{``#1''}

\bibitem{2024shekhar}
S.~Shekhar, W.~Bogaerts, L.~Chrostowski, J.~E. Bowers, M.~Hochberg, R.~Soref, and B.~J. Shastri, \enquote{Roadmapping the next generation of silicon photonics,} {\protect\JournalTitle{Nature Communications}} \textbf{15}, 751 (2024).

\bibitem{2010iqbal}
M.~Iqbal, M.~A. Gleeson, B.~Spaugh, F.~Tybor, W.~G. Gunn, M.~Hochberg, T.~Baehr-Jones, R.~C. Bailey, and L.~C. Gunn, \enquote{Label-free biosensor arrays based on silicon ring resonators and high-speed optical scanning instrumentation,} {\protect\JournalTitle{IEEE Journal of selected topics in quantum electronics}} \textbf{16}, 654--661 (2010).

\bibitem{2022zhang}
X.~Zhang, K.~Kwon, J.~Henriksson, J.~Luo, and M.~C. Wu, \enquote{A large-scale microelectromechanical-systems-based silicon photonics lidar,} {\protect\JournalTitle{Nature}} \textbf{603}, 253--258 (2022).

\bibitem{2019seok}
T.~J. Seok, K.~Kwon, J.~Henriksson, J.~Luo, and M.~C. Wu, \enquote{Wafer-scale silicon photonic switches beyond die size limit,} {\protect\JournalTitle{Optica}} \textbf{6}, 490--494 (2019).

\bibitem{2021shastri}
B.~J. Shastri, A.~N. Tait, T.~Ferreira~de Lima, W.~H. Pernice, H.~Bhaskaran, C.~D. Wright, and P.~R. Prucnal, \enquote{Photonics for artificial intelligence and neuromorphic computing,} {\protect\JournalTitle{Nature Photonics}} \textbf{15}, 102--114 (2021).

\bibitem{2024corsetti}
S.~Corsetti, M.~Notaros, T.~Sneh, A.~Stafford, Z.~A. Page, and J.~Notaros, \enquote{Silicon-photonics-enabled chip-based 3d printer,} {\protect\JournalTitle{Light: Science \& Applications}} \textbf{13}, 132 (2024).

\bibitem{2022ranno}
L.~Ranno, P.~Gupta, K.~Gradkowski, R.~Bernson, D.~Weninger, S.~Serna, A.~M. Agarwal, L.~C. Kimerling, J.~Hu, and P.~OBrien, \enquote{Integrated photonics packaging: Challenges and opportunities,} {\protect\JournalTitle{ACS Photonics}} \textbf{9}, 3467--3485 (2022).

\bibitem{2021minkenburg}
C.~Minkenberg, R.~Krishnaswamy, A.~Zilkie, and D.~Nelson, \enquote{Co-packaged datacenter optics: Opportunities and challenges,} {\protect\JournalTitle{IET Optoelectronics}} \textbf{15}, 77--91 (2021).

\bibitem{2024weninger_patent}
D.~M. Weninger, A.~M. Agarwal, L.~C. Kimerling, and S.~Serna, \enquote{Low-cost, high-performance optoelectronic connectors for integrated circuit packaging,} U.S. Patent Publication No. 2024-0280750 A1 (August 22, 2024).

\bibitem{2024weninger}
D.~Weninger, S.~Serna, L.~Ranno, L.~Kimerling, and A.~Agarwal, \enquote{Low loss chip-to-chip couplers for high-density co-packaged optics,} {\protect\JournalTitle{Advanced Engineering Materials}} \textbf{n/a}, 2402095 (2024).

\bibitem{2006nguyen}
V.~Nguyen, T.~Montalbo, C.~Manolatou, A.~Agarwal, C.-y. Hong, J.~Yasaitis, L.~C. Kimerling, and J.~Michel, \enquote{Silicon-based highly-efficient fiber-to-waveguide coupler for high index contrast systems,} {\protect\JournalTitle{Applied Physics Letters}} \textbf{88}, 081112 (2006).

\bibitem{2007sun}
R.~Sun, V.~Nguyen, A.~Agarwal, C.-y. Hong, J.~Yasaitis, L.~Kimerling, and J.~Michel, \enquote{High performance asymmetric graded index coupler with integrated lens for high index waveguides,} {\protect\JournalTitle{Applied Physics Letters}} \textbf{90}, 201116 (2007).

\bibitem{2010loh}
T.-H. Loh, Q.~Wang, J.~Zhu, K.-T. Ng, Y.-C. Lai, Y.~Huang, and S.-T. Ho, \enquote{Ultra-compact multilayer si/sio2 grin lens mode-size converter for coupling single-mode fiber to si-wire waveguide,} {\protect\JournalTitle{Opt. Express}} \textbf{18}, 21519--21533 (2010).

\bibitem{2016lim}
K.~P. Lim, D.~K.~T. Ng, J.~Pu, Y.~T. Toh, T.~Febiana, K.~Vivek, and Q.~Wang, \enquote{Graded-index thin-film stack for cladding and coupling,} {\protect\JournalTitle{Appl. Opt.}} \textbf{55}, 6752--6756 (2016).

\bibitem{2024xu}
J.~Xu, C.~Guo, Y.~Li, X.~Jin, X.~Quan, and X.~Cheng, \enquote{Graded-index lens based edge coupler with low-loss, broad bandwidth for efficient coupling between silicon waveguide and standard single-mode fiber,} {\protect\JournalTitle{Journal of Lightwave Technology}} \textbf{43}, 701--711 (2025).

\bibitem{2007saleh}
B.~Saleh and M.~Teich, \emph{Fundamentals of photonics} (John Wiley \& Sons Inc., Hoboken, NJ, 2007), pp. 1--40, 2nd ed.

\bibitem{1987bossi}
D.~E. Bossi, J.~M. Hammer, and J.~M. Shaw, \enquote{Optical properties of silicon oxynitride dielectric waveguides,} {\protect\JournalTitle{Appl. Opt.}} \textbf{26}, 609--611 (1987).

\bibitem{2023weninger}
D.~Weninger, S.~Serna, A.~Jain, L.~Kimerling, and A.~Agarwal, \enquote{High density vertical optical interconnects for passive assembly,} {\protect\JournalTitle{Opt. Express}} \textbf{31}, 2816--2832 (2023).

\bibitem{2008sun}
R.~Sun, M.~Beals, A.~Pomerene, J.~Cheng, C.~yin Hong, L.~Kimerling, and J.~Michel, \enquote{Impedance matching vertical optical waveguide couplers for dense high index contrast circuits,} {\protect\JournalTitle{Opt. Express}} \textbf{16}, 11682--11690 (2008).

\bibitem{2021weninger}
D.~M. Weninger, \enquote{Photonic integrated circuit packaging using silicon based optical interconnects,} Master's thesis, Massachusetts Institute of Technology (2021).

\bibitem{2023oxford}
O.~I.~P. Technology, \enquote{Deposition of high quality films by the inductively coupled plasma cvd process,}  (2023). Accessed: 2025-02-28.

\bibitem{mrsi_s_hvm}
Mycronic, \enquote{Mrsi-s-hvm high speed, flexible, 0.5-micron flip-chip die bonder for high volume manufacturing,} \url{https://mrsisystems.com/mrsi-s-hvm/}. Accessed: 2023-12-08.

\bibitem{2017barwicz}
T.~Barwicz, S.~Kamlapurkar, Y.~Martin, R.~L. Bruce, and S.~Engelmann, \enquote{A silicon metamaterial chip-to-chip coupler for photonic flip-chip applications,} in \emph{2017 Optical Fiber Communications Conference and Exhibition (OFC),}  (2017), pp. 1--3.

\bibitem{2012lindenmann}
N.~Lindenmann, G.~Balthasar, D.~Hillerkuss, R.~Schmogrow, M.~Jordan, J.~Leuthold, W.~Freude, and C.~Koos, \enquote{Photonic wire bonding: a novel concept for chip-scale interconnects,} {\protect\JournalTitle{Opt. Express}} \textbf{20}, 17667--17677 (2012).

\bibitem{2018dangel}
R.~{Dangel}, A.~{La Porta}, D.~{Jubin}, F.~{Horst}, N.~{Meier}, M.~{Seifried}, and B.~J. {Offrein}, \enquote{Polymer waveguides enabling scalable low-loss adiabatic optical coupling for silicon photonics,} {\protect\JournalTitle{IEEE Journal of Selected Topics in Quantum Electronics}} \textbf{24}, 1--11 (2018).

\bibitem{2020shaoliang}
S.~Yu, H.~Zuo, X.~Sun, J.~Liu, T.~Gu, and J.~Hu, \enquote{Optical free-form couplers for high-density integrated photonics (offchip): A universal optical interface,} {\protect\JournalTitle{J. Lightwave Technol.}} \textbf{38}, 3358--3365 (2020).

\bibitem{2021englund}
S.~Bandyopadhyay and D.~Englund, \enquote{Alignment-free photonic interconnects,}  (2021).

\bibitem{2021brusberg}
L.~Brusberg, A.~R. Zakharian, {\c S}.~E. Kocaba{\c s}, L.~W. Yeary, J.~R. Grenier, C.~C. Terwilliger, and R.~A. Bellman, \enquote{Glass substrate with integrated waveguides for surface mount photonic packaging,} {\protect\JournalTitle{Journal of Lightwave Technology}} \textbf{39}, 912--919 (2021).

\bibitem{2023brusberg}
L.~Brusberg, J.~R. Grenier, A.~R. Zakharian, L.~W. Yeary, S.-H. Seok, J.-H. Noh, Y.-G. Kim, J.~Matthies, C.~C. Terwilliger, B.~J. Paddock, R.~A. Bellman, D.~W. Levesque, R.~M. Force, C.~G. Sutton, J.~S. Clark, and B.~J. Johnson, \enquote{Glass platform for co-packaged optics,} {\protect\JournalTitle{IEEE Journal of Selected Topics in Quantum Electronics}} \textbf{29}, 1--10 (2023).

\bibitem{2024zhang}
Y.~Zhang, K.~Shang, Y.~Zhang, S.~Li, Y.-C. Lin, and S.~J.~B. Yoo, \enquote{Low-loss and broadband wafer-scale optical interposers for large-scale heterogeneous integration,} {\protect\JournalTitle{Opt. Express}} \textbf{32}, 40--51 (2024).

\bibitem{2024wang}
H.~Wang, Y.~Zuo, X.~Yin, Z.~Chen, Z.~Zhang, F.~Wang, Y.~Hu, X.~Zhang, and C.~Peng, \enquote{Ultralow-loss optical interconnect enabled by topological unidirectional guided resonance,} {\protect\JournalTitle{Science Advances}} \textbf{10}, eadn4372 (2024).

\bibitem{1996bazylenko}
M.~V. Bazylenko and M.~Gross, \enquote{Reactive ion etching of silica structures for integrated optics applications,} {\protect\JournalTitle{Journal of Vacuum Science \& Technology A}} \textbf{14}, 2994--3003 (1996).

\end{thebibliography}
\end{document}